\documentclass[
    ,final            
  ]
  {aipproc}

\layoutstyle{6x9}

\newcommand{\beq}{\begin{equation}}
\newcommand{\eeq}{\end{equation}}
\newcommand{\be}{\begin{eqnarray}}
\newcommand{\ee}{\end{eqnarray}}
\newcommand{\bed}{\begin{displaymath}}
\newcommand{\eed}{\end{displaymath}}
\newcommand{\bea}{\begin{array}}
\newcommand{\eea}{\end{array}}

\begin{document}

\vspace*{-1cm}

\rightline{\sf ADP-06-07/T638, hep-ph/0611058}

\title{Extended Double Lattice BRST, Curci-Ferrari Mass and the Neuberger
  Problem} 


\author{M.~Ghiotti, L.~von~Smekal and A.~G.~Williams}{
  address={Centre for the Subatomic Structure of Matter, University of
  Adelaide, SA 5005, Australia } 
}

\begin{abstract}
We present Extended Double BRST on the lattice and
extend the Neuberger problem to include the ghost/anti-ghost symmetric
formulation of the non-linear covariant Curci-Ferrari (CF) gauges. We then
show how a CF mass regulates the 0/0 indeterminate form of physical
observables, as observed by Neuberger, and discuss the gauge parameter and
mass dependence of the model.
\end{abstract}


\maketitle

\section{Introduction}
In the covariant continuum formulation of gauge theories 
one has to deal with the redundant degrees of freedom due to gauge 
invariance. 
Within the language of local quantum
field theory, 
the machinery for this is based on the
Becchi-Rouet-Stora-Tyutin (BRST) symmetry which can be considered 
the quantum version of local gauge invariance. 
Beyond perturbation theory one faces the famous Gribov ambiguity: the
existence copies of gauge-configurations 
that satisfy the Lorentz condition (or any other local gauge fixing
condition) but are related by gauge transformations, and are thus
physically equivalent. As a result, the usual
definitions of a BRST charge fail to be globally valid.  
A rigorous non-perturbative framework is provided by lattice gauge
theory. Its strength and beauty derives from the fact that
gauge-fixing is not required. However, in order to arrive at a
non-perturbative definition of non-Abelian gauge theories in the
continuum, from a lattice formulation, we need to be able to perform
the continuum limit in a formally watertight way. 
The same ambiguity then shows  
in another form when attempting to fix a gauge
via BRST formulations on the lattice. There it is known as the
Neuberger problem which asserts that the expectation value of any
gauge invariant (and thus physical) observable in a lattice BRST
formulation will always be of the indefinite form 0/0
\cite{Neuberger:1986xz}. 

In this talk we present the ghost/anti-ghost symmetric Curci-Ferrari
gauges with double BRST on the lattice.
We show how Neuberger's argument can be extended to include these
non-linear covariant gauges, and how the indeterminate
form $0/0$ of expectation values is regulated by CF
mass term  \cite{Curci:1976ar} thereby decontracting the
double BRST algebra to its extended version. Finally, we discuss how the
gauge-parameter $\xi$ dependence of the model can be compensated
by adjusting the CF mass with $\xi$.

In pure $SU(N)$ lattice gauge theory, the gauge transformation of link
$U_{ij}$ is defined as $U_{ij}^g = g_i\, U_{ij}\, g^\dagger_j$.
BRST and anti-BRST transformations $s$ and $\bar s$ in
the topological setting do not act on the link variables $U$ directly,
but on the gauge transformations $g_i$ like infinitesimal 
left translations in the gauge group 
with real ghost and anti-ghost Grassmann fields
$c_i^a$, $\bar c_i^a$ as parameters, 
$s g_i  =  c_i  g_i $ and
$\bar s g_i = \bar c_i g_i  $, where 
$ c_i   \equiv c_i^a  X^a $ and  $ \bar c_i   \equiv \bar c_i^a  X^a $.   
For the normalization of the anti-Hermitian generators $X^a$ in
the fundamental representation we use $ \mbox{tr}\, X^aX^b=-\delta^{ab}/2$.
The action of the topological lattice model for gauge fixing {\it a la}
Faddeev-Popov with double BRST invariance can then be written as
\[
S_{\mbox{\tiny CF}} = i s \bar s \Big( V[U^g] + i\xi
\sum_i \mbox{tr}\, \bar c_i c_i \Big)
=\sum_i \Big( i b^a_i F^a_i[U^g] - \frac{i}{2}\bar c^a_i M^a_i[U^g\!,c] + 
\frac{\xi}{2} (b^a_i)^2  + \frac{\xi}{8} 
(\bar c_i \times c_i )^2 
\Big), 
\]
where $V[U^g]= -\sum_i \sum_{j\sim i} \mbox{tr}
 \,  U^g_{ij}= -2\sum_{x,\, \mu} \mbox{Re tr}
 \,  U_{x,\mu}^g$ 
is the gauge fixing functional of covariant gauges which
here assumes the role of a Morse potential on a gauge orbit.
 $F_i^a[U^g] = 0 $ is the gauge-fixing condition and 
$M_i^a[U^g\!,c] = \sum_{j} M_{ij}^{ab} c_j^b$ defines the 
Faddeev-Popov operator of the ghost/anti-ghost symmetric gauges. 
Note the occurrence of quartic ghost self-interactions
$\propto (\bar c_i \times c_i )^2  \equiv (f^{abc} \bar c_i^b c_i^c)^2$ 
which make the Neuberger problem somewhat less obvious in these gauges.
 Details will
 be presented elsewhere.

\section{Regularisation of $0/0$ and $\xi$ independence}

Following Neuberger, we introduce an auxiliary parameter $t$ upfront
the Morse potential, to write the Euclidean partition function used as
the gauge-fixing device, with double BRST,
\beq
\label{partition}
Z_{\mbox{\tiny CF}}(t) = \int d[g, b,\bar c, c]  \;  \exp\Big\{-  i s \bar s
\Big( t V[U^g] + i\xi \sum_i \mbox{tr}\, \bar c_i c_i \Big)
\Big\} \; ,
\eeq
which is independent of $\{U\}$ and $\xi$. Because a derivative
w.r.t.~$t$ produces a BRST-exact operator in the integrand,
it is in fact independent of $t$ also, {\it i.e.}, $Z_{\mbox{\tiny
    CF}}\!\!'(t) = 0$. For $t=0$ on the other hand we find that   
$ Z_{\mbox{\tiny CF}}(0) = 0$; and this is the reason for the
indeterminate form of $0/0$ for all observables first derived for
the standard linear covariant gauges in \cite{Neuberger:1986xz}. The
fact that this conclusion holds also in the ghost/anti-ghost symmetric
formulation with its quartic self-interactions directly relates to the
topological interpretation \cite{Schaden:1998hz} of the Neuberger zero: 
$Z_{\mbox{\tiny CF}}$ can be viewed as the partition function of 
a Witten-type TQFT which computes the  Euler character $\chi$  of the gauge
group. On the lattice the gauge group is a direct product of $SU(N)$'s
per site, and  $Z_{\mbox{\tiny CF}} = \chi(SU(N)^{\#{\rm sites}}) =
\chi(SU(N))^{\#{\rm sites}} = 0^{\# {\rm sites}}$. 
For $t=0$ the action decouples from the link-field
configuration and $Z_{\mbox{\tiny CF}}(0) $, albeit computing the same
topological invariant, has no effect in terms of fixing a gauge. In the present
formulation, $Z_{\mbox{\tiny CF}}(0) $ factorises into independent
Grassmann integrations per site of the quartic term containing the curvature 
of $SU(N)$, each of which computes the vanishing Euler character of
$SU(N)$ via the Gauss-Bonnet theorem \cite{Birmingham:1991ty}.   

As proposed in \cite{Kalloniatis:2005if}, this zero can be
regularised, however, by introducing a Curci-Ferrari mass $m$,
such that the gauge-fixing action $S_{\mbox{\tiny CF}}$ is replaced by 
\beq 
\label{actionCF}
S_{\mbox{\tiny mCF}}(t) = \, i \,( s \bar s-im^2) \, \Big( t V_U[g] + i\xi
\sum_i \mbox{tr}\, \bar c_i c_i \Big).
\eeq
The corresponding partition function $Z_{\mbox{\tiny mCF}}(t)$
no-longer vanishes at $t=0$, and this part in Neuberger's
disastrous conclusion is thus avoided. We have explicitly calculated 
$Z_{\mbox{\tiny mCF}}(0)$, which is polynomial in  $\xi m^4$,
for $SU(2)$ and $SU(3)$. The original zero is obtained for
$m^2 \to 0$ which corresponds to a Wigner-Ionu contraction of the
so-called extended double BRST superalgebra. While a non-vanishing
$m^2$ thereby breaks the nilpotency of BRST and anti-BRST
transformations, which is known to result in a loss of unitarity,
it also serves to regulate the $0/0$ indeterminate form of expectation
values in lattice BRST formulations, and to obtain finite results for
$m^2\to  0 $ via l'Hospital's rule.   

For gauge fixing we need to have $t\not= 0$. The partition function 
$Z_{\mbox{\tiny mCF}}(t)$ of the massive CF model is no-longer
$t$-independent because $s$ and $\bar s$ are no-longer nilpotent and the
simple argument above fails, {\it i.e.},
$Z_{\mbox{\tiny mCF}}\!\!'(t) \not= 0$ for $m^2\not= 0$. However, the
existence of 3 independent parameters $t$, $\xi$ and $m^2$ is an illusion.
A change in $t$ can always be compensated by changing the gauge
parameter $\xi$ and $m^2$. In fact, simple scaling arguments and explicit
calculations show that $Z_{\mbox{\tiny mCF}}$ only depends on  
2 combinations of the 3, we can parametrise  
$Z_{\mbox{\tiny mCF}} = f(t^2/\xi,\xi m^4) \equiv f(x^2,\widehat m^4)$,
where we defined  $x^2 \equiv t^2/\xi $ and $\widehat {m}^4 \equiv \xi
m^4 $. Our explicit calculations for $t=0$ yield $f(0,\widehat m^4)$.
Independence of $t$ then comes together with gauge parameter $\xi$
independence.  To achieve this, we allow
$\widehat m^2 \equiv \widehat {m}^2(x)$ so that $Z_{\mbox{\tiny
    mCF}}=f(x,\widehat {m}^4(x))$. This means that we adjust the CF
mass $\widehat m^2$ with $x$ such that our $x=0$ results remain
unchanged. In particular, we must have 
\beq
\frac{\rm d}{{\rm d}x}
\,Z_{\mbox{\tiny mCF}} = \Big( \frac{\partial}{\partial x} +
\frac{{\rm d} \widehat m^2}{{\rm d}x} \,  \frac{\partial}{\partial
  \widehat m^2} \Big) \,Z_{\mbox{\tiny mCF}} 
= 0 \; ,
\label{adjust}
\eeq
which can be used to determine the derivative of $\widehat m^2(x)$.
This is always possible. The crucial question at this point
is whether it can be done independent of the link configuration  $\{ U\}$.
As our explicit calculations are restricted to $x=0$ we have explicitly
verified that $\widehat m^2{}'(0)$ is  finite and independent of  $\{
U\}$. While this is merely necessary, but not sufficient, it
demonstrates that we can get away from $x=0$, at least
infinitesimally. This is of qualitative importance as a non-zero value
of $x=t/\sqrt{\xi}$, no matter how small, corresponds to a large but
finite $\xi$ at $t=1$ and thus  eliminates the gauge freedom.

\vspace{-.1cm}

\section{Conclusions}

The massive Curci-Ferrari model with extended double BRST symmetry
can be formulated on the lattice without the 0/0 problem. 
The parameter $m^2$ is not interpreted as a physical mass but rather serves 
to meaningfully define a limit $m^2\to 0$ in the spirit of l'Hospital's
rule.  At finite $m^2$ the topological nature of the gauge-fixing
partition function seems lost.  It is possible,
however, to tune the CF mass with the gauge parameter $\xi $ so that the
limit $m^2\to 0$ can be defined along a certain trajectory in
parameter space independent of $\xi $. An
interesting open question might then be the topological
interpretation of the model within the extended double 
BRST superalgebra framework.


\bibliographystyle{aipproc}   

\end{document}